\documentclass[12pt]{article}

\usepackage{amsmath}
\usepackage{graphicx}

\renewcommand{\theequation}{\arabic{section}.\arabic{equation}}
\newcommand{\mys}[1]{\section{#1} \setcounter{equation}{0}}

\newcommand{\myappendix}{\appendix
   \renewcommand{\theequation}{\Alph{section}.\arabic{equation}}
   \vspace{30pt} \noindent {\Large \bf Appendix}}

\newlength{\dummysp}
\newcommand{\diag}{\mathop{{\hbox{diag} \, }}\nolimits}
\newcommand{\tr}{\mathop{{\hbox{Tr} \, }}\nolimits}
\newcommand{\real}{\mathop{{\hbox{Re} \, }}\nolimits}

\newcommand{\stxt}[1]{\mathop{\hbox{{\scriptsize #1}}}\nolimits}
\newcommand{\bbar}[1]{{\overline{#1}}}
\newcommand{\half}{\frac{1}{2}}

\newcommand{\beq}{\begin{eqnarray}}
\newcommand{\eeq}{\end{eqnarray}}
\newcommand{\nnn}{ \nonumber \\ }

\newcommand{\Rbf}{{{\bf R}}}

\newcommand{\e}{{\epsilon}}
\newcommand{\s}{{\sigma}}
\newcommand{\vev}[1]{{\langle #1 \rangle}}
\newcommand{\bigvev}[1]{{\left\langle #1 \right\rangle}}
\newcommand{\ord}[1]{{{\cal O}(#1)}}
\newcommand{\gappeq}{\mathrel{\rlap {\raise.5ex\hbox{$>$}}
{\lower.5ex\hbox{$\sim$}}}}
\newcommand{\lappeq}{\mathrel{\rlap{\raise.5ex\hbox{$<$}}
{\lower.5ex\hbox{$\sim$}}}}
\newcommand{\myref}[1]{(\ref{#1})}

\newcommand{\ben}{\begin{enumerate}}
\newcommand{\een}{\end{enumerate}}

\newcommand{\sqtw}{\sqrt{2}}
\newcommand{\fourth}{\frac{1}{4}}

\newcommand{\psib}{{\bar \psi}}

\newcommand{\bit}{\begin{itemize}}
\newcommand{\eit}{\end{itemize}}

\newcommand{\Cbf}{{\bf C}}

\newcommand{\obf}{{\bf 1}}
\newcommand{\nbf}{{\bf n}}
\newcommand{\mbf}{{\bf m}}
\newcommand{\kbf}{{\bf k}}
\newcommand{\xb}{{\bbar{x}}}
\newcommand{\yb}{{\bbar{y}}}
\newcommand{\ibf}{\boldsymbol{\hat \imath}}
\newcommand{\jbf}{\boldsymbol{\hat \jmath}}
\newcommand{\lbf}{\boldsymbol{\ell}}
\newcommand{\sbf}{\boldsymbol{\s}}

\newcommand{\Ncal}{{\cal N}}

\newcommand{\SSB}{{S_{\stxt{SB}}}}
\newcommand{\hphi}{{\hat \phi}}
\newcommand{\hg}{{\hat g}}
\newcommand{\hmu}{{\hat \mu}}
\newcommand{\hginv}{{{\hat g}^{-1}}}
\newcommand{\muinv}{{{\hat \mu}^{-1}}}
\newcommand{\zb}{{\bbar{z}}}
\newcommand{\xdag}{x^\dagger}
\newcommand{\ydag}{y^\dagger}
\newcommand{\zdag}{z^\dagger}
\newcommand{\hx}{{\hat x}}
\newcommand{\hxd}{{\hat x}^\dagger}
\newcommand{\hxb}{{\hat \xb}}
\newcommand{\hy}{{\hat y}}

\newcommand{\hz}{{\hat z}}
\newcommand{\Dslash}{{\not \hspace{-4pt} D} }
\newcommand{\Ocal}{{\cal O}}
\newcommand{\tmr}{t^{\mu \nu \rho}}
\newcommand{\tmn}{t^{\mu \rho \nu}}

\def\[{\left [}
\def\]{\right ]}
\def\({\left (}
\def\){\right )}

\begin{document}

\begin{titlepage}

\renewcommand{\thefootnote}{\fnsymbol{footnote}}

\hfill May.~21, 2004

\hfill hep-lat/0405021

\vspace{0.45in}

\begin{center}
{\bf \Large Deconstruction, 2d lattice super-Yang-Mills, 
\\ \vskip 10pt and the dynamical lattice spacing}
\end{center}

\vspace{0.15in}

\begin{center}
{\bf \large Joel Giedt\footnote{{\tt giedt@physics.utoronto.ca}}}
\end{center}

\vspace{0.15in}

\begin{center}
{\it University of Toronto \\
60 St. George St., Toronto ON M5S 1A7 Canada}
\end{center}

\vspace{0.15in}

\begin{abstract}
We study expectation values related to the
dynamical lattice spacing that occurs in the recent
2d lattice super-Yang-Mills constructions
of Cohen et al.~[hep-lat/0307012].  The corresponding
observable in the fully-quenched ensemble
would appear to indicate a difficulty with 
the proposed continuum limit.  However, we find that
the same observable in the phase-quenched
ensemble takes a very different, perhaps encouraging,
average value.
Unfortunately, we are not able to obtain results
for the full theory, due to the nearly flat distribution
of the complex phase of the fermion determinant in the 
phase-quenched ensemble.

\end{abstract}

\end{titlepage}

\renewcommand{\thefootnote}{\arabic{footnote}}
\setcounter{footnote}{0}

\mys{Introduction}
In \cite{Cohen:2003qw}, a lattice action has been proposed
by Cohen, Kaplan, Katz and Unsal (CKKU) for the
(4,4) 2d $U(k)$ super-Yang-Mills.\footnote{The
(4,4) 2d $U(k)$ super-Yang-Mills is best defined
as the dimensional reduction of $\Ncal=1$ 6d $U(k)$
super-Yang-Mills; the notation ``(4,4)'' denotes
the number of left and right 2d chirality supercharges.}
The Euclidean target theory action is
\beq
S_{(4,4)} &=& \int d^2 x \; \frac{1}{g_2^2} \tr \[ (Ds_\mu) \cdot (Ds_\mu)
+ \psib_i \Dslash \psi_i + \fourth F \cdot F \right. \nnn
&& \left. 
+ \psib_i [ (s_0 \delta_{ij} + i \gamma_3 {\bf s} \cdot \sbf_{ij}) ,\psi_j] 
- \half [s_\mu, s_\nu]^2 \]
\label{wkkr}
\eeq
where $s_\mu$ ($\mu=0,1,2,3$) are hermitian scalars, $F$ is
the 2d Yang-Mills field strength, and $\psi_i$ ($i=1,2$) are
2d Dirac fermions, all in the adjoint representation
of $U(k)$.  In \cite{Giedt:2003vy}, some aspects
of the fermion determinant were examined, with results
very similar to those in \cite{Giedt:2003ve}.  We will
have occasion to review those results below. 

Here we will
study a rather fundamental aspect of the construction:
the proposed emergence of a ``dynamical'' lattice spacing. 
CKKU make use of an $N \times N$ lattice that contains
link and site fields that are $k \times k$ matrices.
The classical lattice action contains many zero-action
configurations.  CKKU expand the classical lattice action 
about a particular class of zero-action configurations (discussed
below) that are characterized by a parameter $a$.  For this
reason we will refer to such a configuration as an
{\it $a$-configuration.}  In the limit $a \to 0, \; N \to \infty$,
the classical action tends to \myref{wkkr}.
Thus $a$ is interpreted as a lattice spacing.  
However, it is dynamical as it has to
do with a particular background configuration for the lattice
fields.  This strategy is based on the ideas of {\it deconstruction}
\cite{Arkani-Hamed:2001ca,Hill:2000mu}.  Studies
where a partial latticization of 4d supersymmetric theories
has been obtained by this approach include \cite{Giedt:2003xr,Poppitz:2003uz}.

The validity of the semi-classical expansion,
about an $a$-configuration, rests on the assumption
that it gives a good approximation to the behavior of the 
full lattice theory.  But all $a$-configurations are
energetically equivalent.  Furthermore,
there exist other zero action configurations that do
not fall into the $a$-configuration class (shown below).  
Finally, it is important that the fluctuations about
the $a$-configuration be in some sense subdominant.  For these
reasons, CKKU deform the action by adding an $a$-dependent 
potential that favors the $a$-configuration.  The ``continuum
limit'' then includes sending $a \to 0$ in this potential.
Although this deformation breaks the exact lattice supersymmetry,
it is rendered harmless by scaling the relative strength of
the deformation potential to zero in the thermodynamic limit.
For this reason it has been argued by CKKU that the quantum
continuum limit is nothing but the target theory.  Based
on the symmetries of the undeformed theory, it would appear
that the target theory is obtained without
the need for fine-tuning.
For further details, we refer the reader to \cite{Cohen:2003qw},
as well as the articles leading up to 
it \cite{Kaplan:2002wv,Cohen:2003xe}.

The deconstruction method for latticizing a continuum target theory
does not require fermions or supersymmetry.
The bosonic system (i.e., setting all lattice fermions
to zero) contained in the model of
CKKU already has the interesting feature of a deconstructed
lattice Yang-Mills.  The same semi-classical 
arguments yield $U(k)$ Yang-Mills with 4
adjoint scalars; that is, the bosonic part of \myref{wkkr}.
In the present article we will investigate the validity
of the semi-classical argument in both this bosonic
theory and in the full supersymmetric theory.  
We will take into account 
quantum effects, estimating key expectation values
by means of Monte Carlo simulation.

We now summarize the content and results of the rest of the article:
\bit
\item
In the next section we introduce the essential features of
the bosonic part of the CKKU construction that will be needed
for the subsequent discussion.  We then study the
classical minima of the undeformed and deformed actions
in Section \ref{s:cla}.  We show that in the undeformed theory
there is a vast number of zero-action configurations.  We
further show that the deformation leaves only a particular
$a$-configuration with zero action, modulo gauge equivalences.
\item
In Section \ref{s:sim} we outline the methods and results of
our lattice simulation for the bosonic system.  We find
that the expectation values do not agree with the semi-classical
predictions based on energetic arguments.
In Section \ref{intp} we suggest an interpretation
of these results in terms of the analysis of Section \ref{s:cla}.
We find that the simulation results are consistent
with entropic effects dominating the expectation values.
\item
In Section \ref{s:inf} we attempt to include the effects of
fermions.  We show that this is crucial for understanding
the dynamical lattice spacing in the supersymmetric theory, because the
fermion matrix is highly singular for the zero-action
configurations of the undeformed theory.
Difficulties are encountered because of the
complex phase of the fermion determinant.  We present results
for expectation values in the phase-quenched ensemble.
However, we are unable to obtain reliable expectation values
for the full theory, via phase reweighting, 
due to the nearly flat distribution of the complex phase.
\item
In Section \ref{s:con} we make some concluding remarks.  
In the Appendix, we provide a brief review of the $\Ncal=4$
4d super-Yang-Mills moduli space, which arises in the
discussion of Section \ref{s:cla}.
\eit

\mys{Quiver lattice Yang-Mills}
For CKKU, the starting point is the Euclidean
$\Ncal=1$ 6d $U(kN^2)$ super-Yang-Mills.
In our considerations of the bosonic theory, 
the starting point will be just the
Euclidean 6d $U(kN^2)$ Yang-Mills.
In either case, the action is dimensionally 
reduced to 0d to obtain a $U(kN^2)$ matrix model.
The matrix model naturally possesses $SO(6)$ Euclidean invariance.
Next we note $SO(6) \supset SO(2) \times SO(2) \supset 
Z_N \times Z_N$.  CKKU have identified a homomorphic embedding 
of $Z_N \times Z_N$ into the $U(kN^2)$ gauge symmetry group.
Using this, a $Z_N \times Z_N$ orbifold projection is performed to
obtain a $U(k)^{N^2}$ 0d {\it quiver,} or, {\it product group} 
theory.\footnote{Quiver theories were originally studied
many years ago in other contexts \cite{Georgi:au,Halpern:1975yj}.}  
In every respect we follow CKKU, except that in the bosonic
theory we have
set all fermions to zero.  For details of the matrix model and
orbifold projection, we refer the reader to \cite{Cohen:2003qw}.
In the interests of brevity, we will only give the final 
results.  The {\it undeformed} bosonic lattice action is
\beq
S_0 &=& \frac{1}{g^2} \tr \sum_{\nbf} \[
\half (\xdag_{\nbf-\ibf} x_{\nbf-\ibf} - x_{\nbf} \xdag_{\nbf}
+ \ydag_{\nbf-\jbf} y_{\nbf-\jbf} - y_\nbf \ydag_\nbf
+ \zdag_\nbf z_\nbf - z_\nbf \zdag_\nbf )^2 \right. \nnn
&& + 2(x_\nbf y_{\nbf+\ibf} - y_\nbf x_{\nbf+\jbf})
(\ydag_{\nbf+\ibf} \xdag_\nbf - \xdag_{\nbf+\jbf} \ydag_\nbf) \nnn
&& + 2(y_\nbf z_{\nbf+\jbf} - z_\nbf y_\nbf)
(\zdag_{\nbf+\jbf} \ydag_\nbf - \ydag_\nbf \zdag_\nbf) \nnn 
&& \left. + 2(z_\nbf x_\nbf - x_\nbf z_{\nbf+\ibf})
(\xdag_\nbf \zdag_\nbf - \zdag_{\nbf+\ibf} \xdag_\nbf) \]
\label{sbta}
\eeq
The fermionic lattice action will be given in Section \ref{s:inf},
where we will examine the effects of the fermion determinant.
In \myref{sbta}, $x_\mbf, y_\mbf, z_\mbf$ are bosonic lattice
fields that are $k \times k$ unconstrained
complex matrices; $\mbf=(m_1,m_2)$ labels points on an $N \times N$
lattice, and $\ibf=(1,0), \jbf=(0,1)$ are unit vectors.
The $U(k)^{N^2}$ symmetry is nothing but the local $U(k)$
symmetry of the lattice action $S_0$, with link bosons $x_\mbf$ in
the $\ibf$ direction, link bosons $y_\mbf$ in the $\jbf$ direction, and
sites bosons $z_\mbf$, all transforming in the usual manner:
\beq
x_\mbf \to \alpha_\mbf x_\mbf \alpha_{\mbf + \ibf}^\dagger, \qquad
y_\mbf \to \alpha_\mbf y_\mbf \alpha_{\mbf + \jbf}^\dagger, \qquad
z_\mbf \to \alpha_\mbf z_\mbf \alpha^\dagger_\mbf
\label{yuer}
\eeq
Canonical mass dimension 1 is assigned to $x_\mbf, y_\mbf, z_\mbf$,
whereas $g$ has mass dimension 2.

Although $S_0$ is a lattice action that describes a
statistical system with interesting features, it is not in
any obvious way related to a 2d continuum field theory \myref{wkkr}.
As will be explained below, $S_0 \geq 0$ and 
a vast number of nontrivial solutions to $S_0=0$ exist, not all of
which are gauge equivalent.  In fact, the space of minimum
action configurations, or {\it moduli space,} is a multi-dimensional
noncompact manifold with various {\it branches} (classes of
configurations).  For this reason it is difficult to say what a
``continuum limit'' might be; for there exists an infinite number
of energetically equivalent configurations about which to
expand, not all of which are gauge equivalent.

A surprising result---pointed out by CKKU, and based on
ideas from deconstruction---is obtained if one 
expands about the {\it $a$-configuration}
\beq
x_\mbf = \frac{1}{a \sqtw} \obf, \quad
y_\mbf = \frac{1}{a \sqtw} \obf, \quad
z_\mbf = 0, \quad \forall \mbf
\label{ckcf}
\eeq
keeping $g_2=ga$ and $L=Na$ fixed, treating $a$ as small.
(It is easy to see that $S_0=0$ for this configuration.)
That is, we associate $a$ with a lattice spacing (mass
dimensions -1), even though it arises originally from
a specific background field configuration.
In this case, one finds that the classical continuum limit is nothing
but the bosonic part of \myref{wkkr},
which is a variety of 2d $U(k)$ Yang-Mills with adjoint scalars.
In the case of the supersymmetric quiver theory of CKKU,
where fermions are present, one obtains \myref{wkkr}
in full; i.e., (4,4) 2d $U(k)$ super-Yang-Mills.

The trick is how to make the configuration \myref{ckcf}
energetically preferred without destroying all of the
pleasing symmetry properties of the theory.  (This is
particularly true in the supersymmetric case.)  In the
quantum analysis, we must address the more delicate
complication of entropy as well.  CKKU suggest a
deformation of the bosonic action in an effort to stabilize
the theory near the $a$-configuration \myref{ckcf}:
\beq
S_B &=& S_0 + \SSB \\
\SSB &=& \frac{a^2 \mu^2}{2 g^2} \sum_\nbf \tr 
\[ \(x_\nbf \xdag_\nbf -\frac{1}{2a^2}\)^2
+ \(y_\nbf \ydag_\nbf-\frac{1}{2a^2}\)^2 
+ \frac{2}{a^2} z_\nbf \zdag_\nbf \] 
\eeq
Here the strength of the deformation is determined
by the quantity $\mu$, which has mass dimension 1.
It is clear that the configuration \myref{ckcf}
minimizes $\SSB$.  (Other configurations that
minimize $S_0$ and $\SSB$ will be discussed below.)
Unfortunately, in the supersymmetric version of CKKU,
the deformation $\SSB$ breaks the exact supersymmetry of their original
lattice action (hence the subscript ``SB'' = Symmetry Breaking).  For this reason they demand that
the strength of $\SSB$ relative to $S_0$, conveyed by $\mu^2$,
be scaled to zero in the thermodynamic limit.
Thus we are interested in the effects of
the deformation subject to this scaling.

In much of what follows we will specialize to the case
of $U(2)$.  This is merely because it is the simplest
case and the most efficient to simulate.  In this special
case, $x_\mbf, y_\mbf, z_\mbf$ will be unconstrained $2 \times 2$
complex matrices.

\mys{The classical analysis}
\label{s:cla}
Here details are given of the classical analysis of
the minima of the action $S_B$.  In Section \ref{jirr} we
consider the undeformed action $S_0$; then in
Section \ref{wqfe} we consider the modifications induced
by the deformation $\SSB$, which has the effect of
lifting some flat directions in moduli space.
Through understanding this classical picture,
naive expections of what will occur in the quantum
theory, based on energetics, can be formulated.

\subsection{Undeformed theory}
\label{jirr}
Here we neglect $\SSB$ and examine the minima of $S_0$.
Note that \myref{sbta} is a sum of terms of the form
$\tr A A^\dagger$ (the first line involves squares of
hermitian matrices).  Thus $S_0 \geq 0$ with $S_0 = 0$
iff the following equations hold true:
\beq
&& \xdag_{\nbf-\ibf} x_{\nbf-\ibf} - x_{\nbf} \xdag_{\nbf}
+ \ydag_{\nbf-\jbf} y_{\nbf-\jbf} - y_\nbf \ydag_\nbf
+ [\zdag_\nbf, z_\nbf] = 0 
\label{wrse} \\
&& x_\nbf y_{\nbf+\ibf} - y_\nbf x_{\nbf+\jbf} = 
y_\nbf z_{\nbf+\jbf} - z_\nbf y_\nbf = 
z_\nbf x_\nbf - x_\nbf z_{\nbf+\ibf} = 0
\label{wrsf}
\eeq
together with the h.c. of \myref{wrsf}.  The set
of solutions is the moduli space of the undeformed theory.  

\subsubsection{Zeromode branch}
To begin a study of the moduli space,
we isolate the zero momentum modes:  $x_\nbf \equiv x \; \forall
\nbf$, etc.  Then Eqs.~\myref{wrse} and \myref{wrsf} reduce to
\beq
&& [\xdag,x] + [\ydag,y] + [\zdag,z] = 0 \nnn
&& [x,y] = [y,z] = [z,x] = 0
\label{qers}
\eeq
together with the h.c. of the second line of \myref{qers}.  Eqs.~\myref{qers}
may be recognized as nothing but the {\it D-flatness}
and {\it F-flatness} constraints that describe
the moduli space associated with the classical
scalar vacuum of $\Ncal=4$ 4d super-Yang-Mills.\footnote{We
thank Erich Poppitz for pointing this out to us,
as well as the branch of moduli space \myref{kerr} given below.}
The equations are invariant with respect to the global
gauge transformation
\beq
x \to \alpha x \alpha^\dagger, \quad
y \to \alpha y \alpha^\dagger, \quad
z \to \alpha z \alpha^\dagger
\label{oore}
\eeq
Then it is well-known that solutions to \myref{qers}
consist of $x,y,z$ that lie in a Cartan subalgebra
of $U(k)$; the proof is reviewed in Appendix A.  
The global gauge transformations \myref{oore}
allow one to change to a basis where this Cartan subalgebra
has a diagonal realization.  Thus one can think of the
moduli space as the set of all possible diagonal matrices $x,y,z$,
and all global gauge transformations \myref{oore} of this set.

In particular, the zeromode moduli space of the undeformed
$U(2)$ theory is completely described by
\beq
x = x^0 + x^3 \s^3, \quad 
y = y^0 + y^3 \s^3, \quad 
z = z^0 + z^3 \s^3,
\label{uier}
\eeq
with arbitrary complex numbers $x^0,x^3,y^0,y^3,z^0,z^3$,
together with $U(2)$ transformations of these solutions.

Eqs.~\myref{wrse} and \myref{wrsf} also have non-zeromode
solutions.  We do not attempt to present an exhaustive
account of them.  We will merely point out a few such branches
in order to illustrate that the undeformed theory
has a very complicated and large set of $S_0=0$ configurations.
This observation will be relevant to our interpretation
of the simulation results in Section \ref{intp}.
  
\subsubsection{$x_\mbf=y_\mbf=0$  non-zeromode branch}
\label{oops}
We have the very ``large'' branch of moduli space described by
\beq
x_\mbf=y_\mbf=0, \quad z_\mbf = z_\mbf^0 + z_\mbf^3 \s^3, \quad
\forall \mbf
\label{kerr}
\eeq
Again, $z_\mbf^0,z_\mbf^3$ are arbitrary complex numbers.
Furthermore, $z_\mbf$ is a site variable and thus transforms
independently at each site as
\beq
z_\mbf \to \alpha_\mbf z_\mbf \alpha^\dagger_\mbf
\eeq

It can be seen that this branch affords a vast number of solutions to
\myref{wrse} and \myref{wrsf}; there are $N^2$ such
solutions, modulo choices for $z_\mbf^0, z_\mbf^3 \in \Cbf$
and gauge equivalences.  We will argue below that an understanding of
the entropic effects that result from this branch is necessary
to understanding expectation values in the quantum theory---even when
a potential is introduced which gives these configurations
nonvanishing action.

\subsubsection{$z_\nbf =0$ non-zeromode branch}
Another branch in moduli space is the following.  First we
set $z_\nbf =0, \; \forall \nbf$, and introduce Fourier space
variables
\beq
x_\nbf = \frac{1}{N} \sum_\kbf \omega^{\kbf \cdot \nbf} f_\kbf,
\qquad
y_\nbf = \frac{1}{N} \sum_\kbf \omega^{\kbf \cdot \nbf} g_\kbf,
\qquad
\omega = \exp(2\pi i/N)
\eeq
where $\kbf=(k_1,k_2)$ and $k_1, k_2 \in [0,1,\ldots,N-1]$.
Then taking into account $z_\nbf=0$, the conditions \myref{wrse}
and \myref{wrsf} are equivalent to:
\beq
0 &=& \sum_\kbf \( \omega^{\ibf \cdot \lbf} f_\kbf^\dagger f_{\kbf-\lbf}
- f_\kbf f_{\kbf+\lbf}^\dagger
+ \omega^{\jbf \cdot \lbf} g_\kbf^\dagger g_{\kbf-\lbf}
- g_\kbf g_{\kbf+\lbf}^\dagger \) \nnn
0 &=& \sum_\kbf \( \omega^{-\ibf \cdot (\lbf + \kbf)} f_\kbf g_{-\kbf-\lbf}
- \omega^{\jbf \cdot \kbf} g_{-\kbf-\lbf} f_\kbf \)
\label{klsf}
\eeq
for all $\lbf=(\ell_1,\ell_2)$ and $\ell_1, \ell_2 \in [0,1,\ldots,N-1]$.
Next we turn off all modes except one for both $f_\kbf$ and $g_\kbf$:
\beq
f_\kbf = \delta_{\kbf,\kbf'} f_{\kbf'}, \qquad
g_\kbf = \delta_{\kbf,-\kbf'} g_{-\kbf'}
\eeq
Here and below, {\it no} sum over $\kbf'$ is implied.
When substituted into \myref{klsf}, only 2 nontrivial
conditions survive:
\beq
0 = [f_{\kbf'}^\dagger, f_{\kbf'}]
+ [g_{-\kbf'}^\dagger, g_{-\kbf'}], \qquad
0 = f_{\kbf'} g_{-\kbf'} 
- \omega^{(\ibf+\jbf) \cdot \kbf'} g_{-\kbf'} f_{\kbf'}
\label{klsg}
\eeq
For the $U(2)$ case, we find that solutions exist if
$\omega^{(\ibf+\jbf)\cdot \kbf'} = \pm 1$.  We already know from the
zeromode considerations that for $\omega^{(\ibf+\jbf)\cdot \kbf'} = 1$
we have solutions for $f_{\kbf'}, g_{-\kbf'}$
diagonal matrices.  In the case of $\omega^{(\ibf+\jbf)\cdot \kbf'} = -1$
it is easy to see that there are solutions, say, of the form
\beq
f_{\kbf'} = z_f \s^3, \quad g_{-\kbf'} = z_g (\s^1 + b \s^2),
\quad z_f, z_g \in \Cbf, \quad b \in \Rbf
\label{ppor}
\eeq
There are many values of $\kbf'$ for
which $\omega^{(\ibf+\jbf)\cdot \kbf'} = \pm 1$.  
For $N$ even these are
\beq
k'_1 + k'_2 = 0, \frac{N}{2}, N, \frac{3N}{2}
\eeq
For $N$ odd, $k'_1 + k'_2 = 0, N$ are allowed and in the cases
where
\beq
k'_1 + k'_2 = \frac{N \pm 1}{2}, \frac{3 (N\pm 1)}{2}
\eeq
\myref{ppor} yield approximate solutions to \myref{klsg}, with an error
of order $1/N$.  Thus in the $N \to \infty$ limit the number 
of $S_0=0$ configurations in this class is vast; in fact, it
is easy to check that the number of such configurations is
approximately $2N$, modulo gauge equivalences and
various choices for the constants in \myref{ppor}.

\subsection{Deformed theory}
\label{wqfe}
Now we consider the supersymmetry breaking
deformation $\SSB$ introduced by CKKU.  
To see its effect it is handy to rewrite
the quantities that appear in it.  Recall that $x_\mbf$ is a
complex $2 \times 2$ matrix.  Dropping the subscript, we
can always define
\beq
x=x^0 + x^a \s^a, \qquad x^\dagger = \xb^0 + \xb^a \s^a
\eeq
Then it is straightforward to work out ($\mu=0,\ldots,3$)
\beq
x x^\dagger = x^\mu \xb^\mu + (x^0 \xb^c + \xb^0 x^c
+ i x^a \xb^b \e^{abc} ) \s^c \equiv \phi^{x,0} + \phi^{x,c} \s^c
\equiv \phi^x
\eeq
Note that $\phi^{x,\mu}$ are real, and
that $\phi^{x,0}$ is positive definite.  With
similar definitions for $\phi^y, \phi^z$, the CKKU deformation is
\beq
\SSB &=& \frac{a^2 \mu^2}{2 g^2} \sum_\mbf \tr \[ \(\phi_\mbf^x-\frac{1}{2a^2}\)^2
+ \(\phi_\mbf^y-\frac{1}{2a^2}\)^2 + \frac{2}{a^2} \phi_\mbf^z \] \nnn
&=& \frac{a^2 \mu^2}{g^2} \sum_\mbf \[ \( \phi_\mbf^{x,0} - \frac{1}{2a^2}\)^2
+ \( \phi_\mbf^{y,0} - \frac{1}{2a^2}\)^2 
+ \frac{2}{a^2} \phi_\mbf^{z,0} \right. \nnn 
&& \quad \left. + \sum_a \[ (\phi_\mbf^{x,a})^2 + (\phi_\mbf^{y,a})^2 \] \]
\eeq
It can be seen that the deformation drives $\phi_\mbf^{x,a},\phi_\mbf^{y,a},
\phi_\mbf^{z,0}$ toward the origin, and $\phi_\mbf^{x,0}, \phi_\mbf^{y,0}$ 
toward $1/2a^2$.  When $\phi_\mbf^{z,0}=0$, 
it is easy to see that $\phi_\mbf^{z,a}=0$ identically.

To continue the analysis, 
it is convenient to rescale to dimensionless quantities using
the parameter $a$:
\beq
\hat g = g a^2, \quad \hat \mu = \mu a, \quad \hat \phi_\mbf^x = a^2 \phi_\mbf^x,
\quad \hx_\mbf = a x_\mbf, \quad {\rm etc.}
\label{krkr}
\eeq
Then
\beq
\SSB &=& \frac{{\hat \mu}^2}{{\hat g}^2} 
\sum_\mbf \[ \( \hphi_\mbf^{x,0} - \frac{1}{2}\)^2
+ \( \hphi_\mbf^{y,0} - \frac{1}{2}\)^2 
+ 2 \hphi_\mbf^{z,0}
+ \sum_a \[ (\hphi_\mbf^{x,a})^2 
+ (\hphi_\mbf^{y,a})^2 \] \]
\label{pool}
\eeq
For any value of the lattice spacing $a$, the minimum of $\SSB$ is
obtained iff
\beq
\hphi_\mbf^{x,0} = \hphi_\mbf^{y,0}= \half, \quad
\hphi_\mbf^{z,0} = \hphi_\mbf^{x,a} =\hphi_\mbf^{y,a} = 0,
\quad \forall \mbf
\label{jjus}
\eeq

The conditions involving $x_\mbf$ are just
\beq
\hx_\mbf^\mu \hxb_\mbf^\mu = \half, \qquad
\hx_\mbf^0 \hxb_\mbf^c + \hxb_\mbf^0 \hx_\mbf^c 
+ i \hx_\mbf^a \hxb_\mbf^b \e^{abc} = 0
\label{piwr}
\eeq
Let us examine what additional constraint this places
on classical solutions to $S=0$, beyond the restrictions
of the undeformed theory.  

First we note that neither of the non-zeromode branches
discussed in Section \ref{jirr} above are minima of
$\SSB$.  Thus we pass on to the zeromode configurations \myref{uier}.
Eqs.~\myref{piwr} then imply
that \myref{uier} is restricted to the form
\beq
\hx = \frac{e^{i \gamma_x}}{\sqtw} \diag \( e^{i \varphi_x},e^{-i \varphi_x} \)
\label{xsom}
\eeq
and global gauge transformations of this.
That is, $\hx$ is restricted to be an element of the 
maximal abelian subgroup $U(1)^2$ of $U(2)$, up to an
overall factor of $1/\sqtw$.  Similarly, we have for $\hy$,
\beq
\hy = \frac{e^{i \gamma_y}}{\sqtw} \diag \( e^{i \varphi_y},e^{-i \varphi_y} \)
\label{ysom}
\eeq
Finally, $\hphi_\mbf^{z,0} = 0$ implies $z_\mbf^\mu \zb_\mbf^\mu = 0$,
which has the unique solution $z_\mbf=0, \; \forall \mbf$.

Apart from global obstructions that are essentially Polyakov loops
in the $\ibf$ or $\jbf$ directions, the configuration \myref{xsom}
and \myref{ysom} can be gauged away.  It is straightforward to
verify that the required gauge transformation is \myref{yuer} with
\beq
\alpha_{m_1,m_2} = e^{i (m_1 \gamma_x + m_2 \gamma_y)} 
\times \diag \( e^{i(m_1 \varphi_x + m_2 \varphi_y)},
 e^{- i(m_1 \varphi_x + m_2 \varphi_y)} \)
\eeq
This sets all $\hx_\mbf,\hy_\mbf$ to unity except at the ``boundaries'':
\beq
\hx_{N,m_2} &=& \frac{e^{i N \gamma_x}}{\sqtw} 
\diag \( e^{i N \varphi_x},e^{-i N\varphi_x} \), \quad \forall m_2 \nnn
\hy_{m_1,N} &=& \frac{e^{i N \gamma_y}}{\sqtw} 
\diag \( e^{i N \varphi_y},e^{-i N\varphi_y} \), \quad \forall m_1
\eeq
For most purposes, we do not expect such vacua to distinguish
themselves from the trivial vacua in the thermodynamic limit.
In any case, global features such as these are typical of classical
vacua of other lattice Yang-Mills formulations, such as the Wilson
action.  In our simulation study we will avoid this issue by
restricting our attention to the expectation value of
a quantity that is independent of these angles.

\mys{Simulation of the bosonic theory}
\label{s:sim}
The emergence of the effective lattice theory relies upon
the assumption that fluctuations about this classical minimum
are small, and that the equations \myref{jjus}
are a good approximation to the corresponding expectation values
in the quantum theory.  For this reason we study
\beq
\vev{\hphi_\mbf^{x,0}} = \vev{\hx_\mbf^\mu \hxb_\mbf^\mu}
= \bigvev{ \half \tr (\hx_\mbf \hxd_\mbf) }
\label{iwre}
\eeq
in our simulations, and compare the
expectation values to the classical prediction \myref{jjus}.

\subsection{Scaling}
We study \myref{iwre} along a naive scaling trajectory:
\beq
g_2 = a^{-1} \hat g(a) = \mbox{fixed}
\label{yuie}
\eeq
That is, we hold the bare coupling in physical units, $g_2$, fixed;
this is equivalent to neglecting its anomalous dimension.  The dimensionless
bare coupling $\hat g$ is then a function of $a$ that vanishes
linearly with $a$ as the UV cutoff is removed.  

With regard to $\hmu$ we follow the instructions of CKKU:  
we send the dimensionless coefficient $\hat \mu$ of the deformation $\SSB$ to
zero as $1/N$ while increasing $N$.  
\beq
\hat \mu^{-1} = c N,  \qquad c = \ord{1}
\label{pwer}
\eeq
This is equivalent to scaling 
$\mu = 1/cL$, where $c$ is a constant and $L=Na$ is the
extent of the system.  

In the rescaled variables \myref{krkr},
the coefficient of the undeformed action is $1/\hg^2$,
whereas the coefficient of the deformation is $\hmu^2/ \hg^2$,
as can be seen from \myref{pool}.  Thus it is that
the relative strength of the deformation vanishes
in the thermodynamic limit, when \myref{pwer} is imposed.

We perform these scalings for a sequence of decreasing
values of $a$.  That is, we
study the thermodynamic limit for fixed values of $a$.  We then
extrapolate toward $a=0$ to obtain the continuum limit.

The physical length scales are set by 
$g_2^{-1}$ and the system size $L=Na$.
To keep discretization effects to a minimum we would like to
take $g_2^{-1} \gg a$.  Equivalently, $\hg^{-1} \gg 1$.  On the other
hand we are most interested in what happens at large or infinite
volume.  To render finite volume effects negligible would
require $g_2^{-1} \ll L$.  Equivalently, $\hg^{-1} \ll N$.
In the simulations we study the system for 
various choices of $\hg^{-1}$ and $N$.
We extrapolate to the regime $1 \ll \hg^{-1} \ll N$, but
often violate the bounds $1 \leq \hg^{-1} \leq N$ for specific
points where measurements are taken.  The reason for this
is that data outside the optimal window $1 \ll \hg^{-1} \ll N$
is informative to the extrapolation.

\subsection{Sampling procedure}
We update the system using a multi-hit Metropolis algorithm.
We attempt to update a single site or link
field 10 times before moving to the next, with an acceptance
rate of approximately 50 percent for a single hit.  We find
that this minimizes autocorrelations while maintaining program
efficiency.  We have examined autocorrelations and the
dependence of our observables on initial conditions.  
These studies have led us to make
500 themalization sweeps after a random initialization,
and 100 updating sweeps between each sample.  1/2 to 2 percent
standard errors result from accumulating 1000 samples at
each data point.

\subsection{Results}
\begin{figure}
\begin{center}
\includegraphics[height=5.0in,width=5.0in,angle=90]{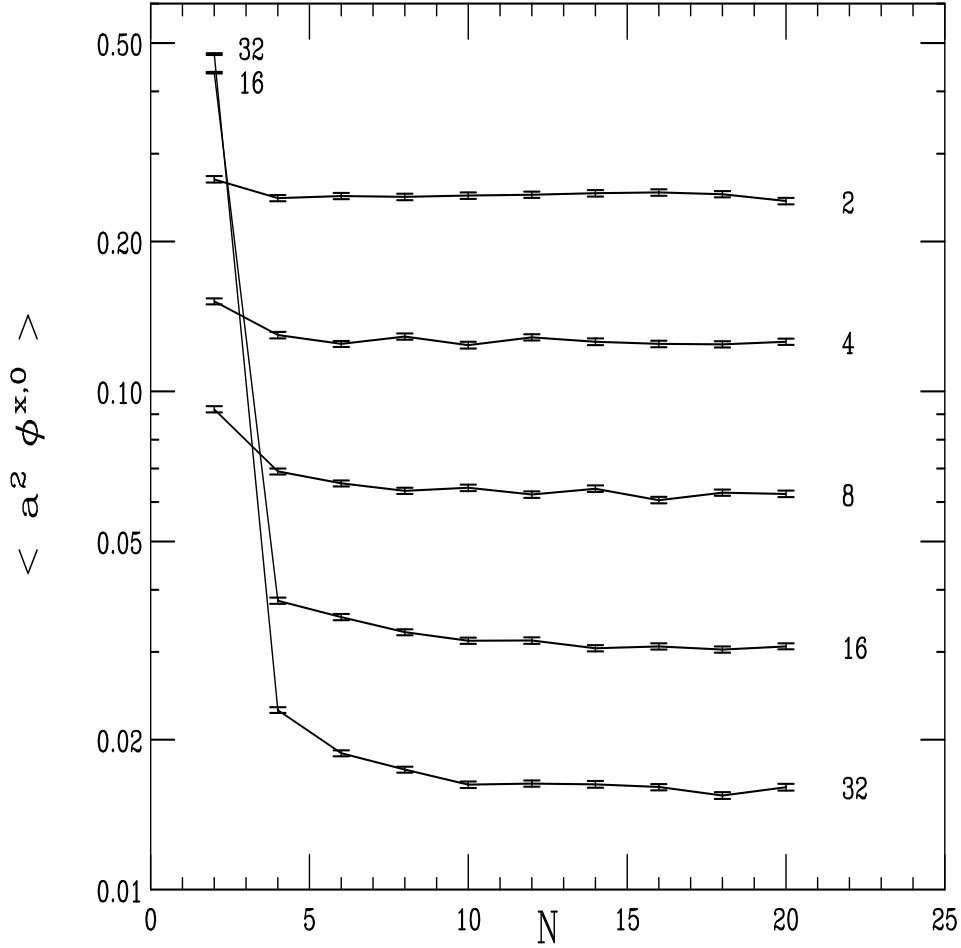}
\end{center}
\caption{Trajectories of fixed lattice spacing,
increasing volume, with data connected by
lines to guide the eye.  Each line is marked
by the corresponding value of $\hg^{-1}$.
A doubling of $\hg^{-1}$ corresponds
to a halving of the lattice spacing.}
\label{f1}
\end{figure}

In Fig.~\ref{f1} we show $\vev{\hphi_\mbf^{x,0}}$ [cf.~\myref{iwre}]
as a function of $N$
for various values of $\hg^{-1}$, having set $c=1$
in \myref{pwer}.  Doubling $\hg^{-1}$
is equated with halving the lattice spacing, according to \myref{yuie}.
It can be seen that $\vev{\hphi_\mbf^{x,0}}$ tends toward
smaller values as the lattice regulator is removed, contrary to
the classical expectations \myref{jjus}.  At large enough values of $N$
the curves flatten out to a constant value. 

\begin{figure}
\begin{center}
\includegraphics[height=5.0in,width=2.5in,angle=90]{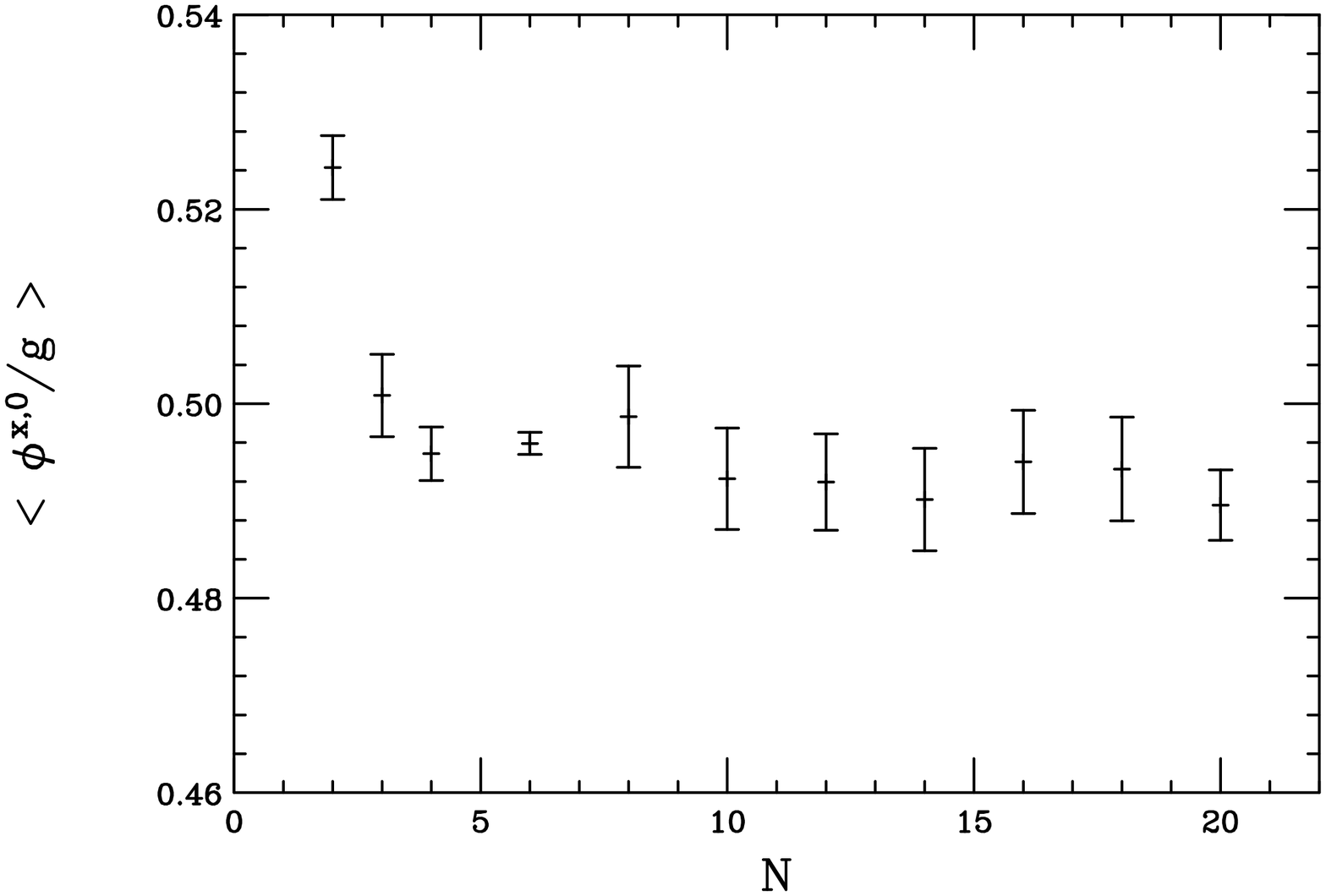}
\end{center}
\caption{Undeformed theory results.  Error bars differ
in size due to differing numbers of samples.
\label{f2}}
\end{figure}

To understand this behavior, we first note that we are computing
an expectation value that is already nonvanishing in the undeformed
theory ($\mu \equiv 0$).  The undeformed theory expectation values 
\beq
\vev{g^{-1} \phi_\mbf^{x,0}}=
\vev{\hg^{-1} \hphi_\mbf^{x,0}}
\label{uter}
\eeq
for various values of $N$ are shown in Fig.~\ref{f2}.  The
rescaling by $g^{-1} = a^2 \hg^{-1}$ is useful because it corresponds to
removing $g$ from the undeformed action $S_0$ 
by a rescaling of the lattice variables [cf.~\myref{sbta}]; 
this amounts to studying the undeformed theory in units of $\sqrt{g}$.
It can be seen from Fig.~\ref{f2} that \myref{uter} is far
from zero, and is rather insensitive to $N$.  The undeformed
theory only contains two length scales, $1/\sqrt{g}$ and $N/\sqrt{g}$.
For large $N$, it is not surprising that the {\it local} expectation
value \myref{uter} is insensitive to the long 
distance scale $N/\sqrt{g}$.  Rather,
it is determined by the short distance scale $1/\sqrt{g}$.
The deformed theory retains this short distance scale.  Once the
system size is much larger than this scale, the local expectation
value becomes independent of the system volume; this is particularly
true because the relative strength of the deformation is being scaled
to zero [cf.~\myref{pwer}].

In Table \ref{tb2} we show the large $N$
expectation values \myref{uter} in the deformed theory
as well as those of the undeformed theory.
Here, the values for $N=16, 18, 20$
were averaged for each $\hginv$, which should provide
a good estimate of the asymptotic value, as can be
seen in Fig.~\ref{f1}.  The error was estimated
based on the maximum deviation from this mean, among
the three data points, taking into account the $1\s$ error
estimates that have been represented in the figure
by error bars.  Table \ref{tb2} shows that the large $N$
expectation values \myref{uter} {\it in the deformed theory}
are (up to statistical errors) the same as those of the 
{\it undeformed theory.}

\begin{table}
\begin{center}
\begin{tabular}{ccc}
$\hginv$ & $\vev{\hat \phi^{x,0}}$ & $\vev{\hg^{-1} \hphi^{x,0}}$ \\ \hline
2 & 0.2469(94) & 0.494(19) \\
4 & 0.1249(27) & 0.500(11) \\
8 & 0.0618(22) & 0.494(18) \\
16 & 0.03064(75) & 0.490(12) \\
32 & 0.01586(64) & 0.507(21) \\ \hline
1 ($\mu \equiv 0$) & & 0.496(22) \\ \hline
\end{tabular}
\caption{Large $N$ asymptotic values for the $\muinv=N$ trajectories.
For comparison, the result for the undeformed ($\mu \equiv 0$) expectation value
is shown in the bottom line.  Estimated errors in the last 2 digits
are shown in parentheses.
\label{tb2}}
\end{center}
\end{table}

Incidentally, one might worry that due to the nontrivial moduli
space of the undeformed theory, its partition function (for the
purely bosonic theory) is not well-defined.  Certainly this
is the case for case for the 0d reduction of $d \leq 4$ $SU(2)$ pure
Yang-Mills; however, this is not the case---in spite of noncompact
flat directions---for the 0d reduction of $d \geq 5$ $SU(2)$ pure
Yang-Mills; see Eq.~(24) of \cite{Krauth:1998xh}.  This is
because, properly speaking, the classical moduli space is a set
of measure zero in the field integration of the partition function.
For $d \geq 5$, ``entropic effects of the measure...overwhelm
the possible divergences'' \cite{Krauth:1998xh}.  
We are not aware of any exact result
showing that the partition function of the CKKU quiver system
obtained from the $6d \to 0d$ $SU(kN^2)$ pure Yang-Mills
is divergent.\footnote{The diagonal $U(1)$ from $U(kN^2) = U(1) 
\times SU(kN^2)$ decouples; its (divergent) contribution
to the partition function trivially factors out.}
In our view it is unlikely, just from the result
that $6d \to 0d$ $SU(2)$ pure Yang-Mills has a finite partition
function.  Furthermore, if there was a problem with the partition
function of the undeformed theory, we would have expected an
uncontrolled dispersion in our observables when we attempted
to measure them in the undeformed theory.  This is because
flat directions not suppressed by entropic effects would allow
for the Monte Carlo simulation to wander wildly 
throughout configuration space,
leading to results that would not converge to reliable average
values.  That this did not occur is further evidence that
the partition function is well-defined.  Finally, the deformation $S_{SB}$
can be regarded as a regulator of any possible divergence associated
with the noncompact flat directions.  That we obtain stable, identical
results as the deformation is removed provides further evidence that
our expectation values for the undeformed theory are reliable.

\mys{Interpretation}
\label{intp}
In our simulations of the purely bosonic 
system, we have observed that under the 
scaling \myref{pwer}, the deformation becomes ineffective
at changing the expectation value \myref{uter}, or
equivalently \myref{iwre}, away from the value that
would be obtained in the undeformed theory.  What has
happened is that the flat directions that were
lifted by the deformation are becoming flat all over
again as $N \to \infty$.  More precisely, the deformation
is proportional to $1/g_2^2 L^2$, and in the thermodynamic
limit this quantity vanishes.  Whereas
the configurations of the moduli space of the undeformed theory
that were lifted by $\SSB$ cost energy at finite $N$, 
the number of such configurations is
becoming vast due to the approximate flatness
in those directions.  For $N \gappeq \hg^{-1/2}$,
the entropy of these configurations wins out over
the energy arguments that prefer \myref{jjus}.

One lifted region of the $S_0$ moduli space 
where this is particularly clear is the branch \myref{kerr}.
The action for such configurations is (setting $c=1$)
\beq
S_B=\frac{1}{2 \hg^2} + \frac{2}{\hg^2 N^2} \sum_\mbf \(
|\hz_\mbf^0|^2 + |\hz_\mbf^3|^2 \)
= \frac{N^2}{2 g_2^2 L^2} + \frac{2}{g_2^2 L^2} \sum_\mbf \(
|\hz_\mbf^0|^2 + |\hz_\mbf^3|^2 \)
\eeq
Integrating $\exp(-S_B)$ over all $\hz_\mbf^{0,3}$ we obtain
\beq
\(\frac{\pi}{2}\)^{2N^2} 
\exp\[ \frac{N^2}{2 g_2^2 L^2} \( -1 + 4 g_2^2 L^2 \ln (g_2^2 L^2) \) \]
\eeq
Note that for large system size $g_2 L \gg 1$.  Thus, the positive
(entropic) term under the exponential wins out over the
negative (energetic) term by a large margin.  It would seem that
the weight of these configurations increases exponentially
as we increase $N$ while holding $g_2 L$ fixed; that is,
in the continuum limit.

Indeed the observed behavior summarized in Table~\ref{tb2}
is that in the deformed theory
\beq 
\bigvev{ \half \tr (\hx_\mbf \hxd_\mbf) } \approx \frac{g_2 L}{2N}
= \half g_2 a
\eeq
Thus the simulations likewise indicate that configurations
with $\hx_\mbf=0$ are dominating as we increase $N$ while
holding $g_2 L$ fixed, which is nothing but the continuum
limit.  By symmetry, the same also holds for $\hy_\mbf$.

\mys{Including fermions}
\label{s:inf}
While the above results and arguments for the non-supersymmetric
system are interesting in their own right, it is essential to
include the fermions if we are to draw conclusions for the
supersymmetric system.

Unfortunately, the present system suffers from a complex
fermion determinant \cite{Giedt:2003vy}.  Thus the integration
measure $\det M e^{-S_B}$ is not positive semi-definite
and does not provide a satisfactory probability measure
for Monte Carlo simulations.  If we
factor out the phase $e^{i \alpha} = \det M / |\det M|$ 
and use $|\det M| e^{-S_B}$
instead, then we will generate the {\it phase-quenched (p.q.)}
ensemble of boson configurations.  Expectation values
of an operator $\Ocal$ in the full theory are formally
related to those in the phase-quenched theory by the
{\it reweighting} identity
\beq
\vev{\Ocal} = \vev{e^{i \alpha} \Ocal}_{p.q.} /
\vev{e^{i \alpha}}_{p.q.}
\label{tslr}
\eeq

In some lattice systems, the distribution of $\alpha$ in the 
phase-quenched ensemble is sharply peaked. 
The phase-quenched quantities in the ratio exist 
and can be measured reliably.  For example, this is the case in the 
4d $U(1)_L \times U(1)_R$ symmetric Yukawa model of \cite{Munster:1992jq}.
By contrast, the results of \cite{Giedt:2003vy} for the
CKKU system are not encouraging.  The distribution of $\alpha$
(denoted $\phi$ in that work) in the phase-quenched ensemble 
was found to be essentially flat.  This leads to approximate
cancellations when we attempt to compute reweighted quantities
contained in the numerator and denominator of \myref{tslr}.
More details will be given below.  Here we merely emphasize
that we do not expect the strategy implied by \myref{tslr}
to succeed, in practice, for the present system.  Our chief
result will thus be to quantify the degree to which it fails.

As will be seen below in Sec.~\ref{dms}, $\det M$ vanishes for
the $S_0=0$ configurations that were discussed above.
Thus the configurations that have the most weight in the
fully-quenched ensemble are configurations that have vanishing
weight in the phase-quenched ensemble.  (Further quantification
of this statement is given in Sec.~\ref{erew}.)
For this reason we find it interesting to study the quantity 
$\vev{\hphi_\mbf^{x,0}}_{p.q.}$, in Sec.~\ref{oope}.  Certainly this
bears a closer resemblance to the expection value in the
full theory, in that the suppression due to very small
$|\det M|$ is taken into account for configurations in
the neighborhood of the undeformed moduli space.  In Sec.~\ref{pooe},
we return to the matter of phase reweighting.

\subsection{Fermion action}
\label{uies}
First we review the fermion action in the notation of \cite{Giedt:2003vy}.
The fermion action can be written in the form
\beq
S_F = - \frac{1}{g^2}
\( \psi_{1,\mbf}^\mu ~,~ \psi_{2,\mbf}^\mu ~,~
\psi_{3,\mbf}^\mu ~,~ \chi_\mbf^\mu \)
\cdot M_{\mbf \nbf}^{\mu \rho} \cdot
\begin{pmatrix} \xi_{1, \nbf}^\rho \cr
\xi_{2, \nbf}^\rho \cr \xi_{3, \nbf}^\rho
\cr \lambda_\nbf^\rho \cr \end{pmatrix}
\label{haht}
\eeq
The elements of the fermion matrix are:
\beq
(M_{\mbf \nbf}^{\mu \rho})_{1,1} &=&
(M_{\mbf \nbf}^{\mu \rho})_{2,2} \; = \;
(M_{\mbf \nbf}^{\mu \rho})_{3,3} \; = \;
(M_{\mbf \nbf}^{\mu \rho})_{4,4} \; = \; 0,
\nnn
(M_{\mbf \nbf}^{\mu \rho})_{1,2} &=&
- \tmr_{\mbf,\nbf} z_{\nbf+\ibf}^\nu + \tmn_{\mbf,\nbf} z_\nbf^\nu ,
\qquad
(M_{\mbf \nbf}^{\mu \rho})_{1,3} =
\tmr_{\mbf,\nbf} y_{\nbf+\ibf}^\nu - \tmn_{\mbf,\nbf+\jbf} y_\nbf^\nu ,
\nnn
(M_{\mbf \nbf}^{\mu \rho})_{1,4} &=&
\tmr_{\mbf,\nbf} \xb_{\nbf}^\nu - \tmn_{\mbf,\nbf-\ibf} \xb_\mbf^\nu ,
\qquad
(M_{\mbf \nbf}^{\mu \rho})_{2,1} =
\tmr_{\mbf,\nbf} z_{\nbf+\jbf}^\nu - \tmn_{\mbf,\nbf} z_\nbf^\nu ,
\nnn
(M_{\mbf \nbf}^{\mu \rho})_{2,3} &=&
- \tmr_{\mbf,\nbf} x_{\nbf+\jbf}^\nu + \tmn_{\mbf,\nbf+\ibf} x_\nbf^\nu ,
\qquad
(M_{\mbf \nbf}^{\mu \rho})_{2,4} =
\tmr_{\mbf,\nbf} \yb_{\nbf}^\nu - \tmn_{\mbf,\nbf-\jbf} \yb_\mbf^\nu ,
\nnn
(M_{\mbf \nbf}^{\mu \rho})_{3,1} &=&
- \tmr_{\mbf,\nbf} y_{\nbf}^\nu + \tmn_{\mbf,\nbf+\jbf} y_\nbf^\nu ,
\qquad
(M_{\mbf \nbf}^{\mu \rho})_{3,2} =
\tmr_{\mbf,\nbf} x_{\nbf}^\nu - \tmn_{\mbf,\nbf+\ibf} x_\nbf^\nu ,
\nnn
(M_{\mbf \nbf}^{\mu \rho})_{3,4} &=&
\tmr_{\mbf,\nbf} \zb_{\nbf}^\nu - \tmn_{\mbf,\nbf} \zb_\nbf^\nu ,
\qquad
(M_{\mbf \nbf}^{\mu \rho})_{4,1} =
- \tmr_{\mbf,\nbf} \xb_{\nbf+\jbf}^\nu
+ \tmn_{\mbf,\nbf-\ibf} \xb_\mbf^\nu ,
\nnn
(M_{\mbf \nbf}^{\mu \rho})_{4,2} &=&
- \tmr_{\mbf,\nbf} \yb_{\nbf+\ibf}^\nu
+ \tmn_{\mbf,\nbf-\jbf} \yb_\mbf^\nu ,
\qquad
(M_{\mbf \nbf}^{\mu \rho})_{4,3} =
- \tmr_{\mbf,\nbf} \zb_{\nbf+\ibf+\jbf}^\nu
+ \tmn_{\mbf,\nbf} \zb_\nbf^\nu .
\eeq
\beq
&& T^\mu = ( \obf_2, \s^a ), \qquad
\tr (T^\mu T^\nu T^\rho) = 2 t^{\mu \nu \rho} \qquad
\Rightarrow \nnn
&& t^{000}=1, \qquad t^{\underline{a00}} = 0, \qquad
t^{\underline{ab0}} = \delta^{ab}, \qquad t^{abc} = i\e^{abc} \nnn
&& t_{\mbf, \nbf}^{\mu \nu \rho} = \delta_{\mbf,\nbf} t^{\mu \nu \rho}
\eeq

This matrix has two ever-present fermion zeromodes.
The first is associated with left multiplication
(the transpose ``T'' merely indicates a row vector,
for consistency with \myref{haht}):
\beq
(\xi_{1, \nbf}^\rho , \xi_{2, \nbf}^\rho , \xi_{3, \nbf}^\rho ,
\lambda_\nbf^\rho )^T
= (0 , 0 , 0 , \lambda \delta^{\rho 0} )^T \quad
\forall \quad \nbf
\label{zv1}
\eeq
The second is associated with right multiplication:
\beq
\( \psi_{1,\mbf}^\mu ~,~ \psi_{2,\mbf}^\mu ~,~
\psi_{3,\mbf}^\mu ~,~ \chi_\mbf^\mu \)
= \( 0 ~,~ 0 ~,~
\psi \delta^{0 \mu} ~,~ 0 \) \quad \forall \quad \mbf
\label{zv2}
\eeq
Because the matrix $M$ is not hermitian,
it is diagonalized by $M \to D = U M V$
with $U$ and $V$ independent unitary matrices.
When this is done, the diagonal matrix $D$
always has 2 zeros on the diagonal, one each from
\myref{zv1} and \myref{zv2}.

The zeromode eigenvalues of the daughter theory
can be factored out following the method used
in \cite{Giedt:2003vy,Giedt:2003ve}.  We
deform the fermion matrix appearing in \myref{haht} according to
\beq
M \to M_\e \equiv M + \e {\bf 1}_{N_f}
\eeq
where $N_f = 16 N^2$ is the dimensionality of the fermion matrix
and $\e \ll 1$ is a deformation parameter that we will
eventually take to zero.  We factor out the zero mode
through the definition
\beq
&& \hat M_0 = \lim_{\e \to 0^+} \hat M_\e, \quad
\hat M_\e \equiv \e^{-2/N_f} M_\e  \; \Rightarrow \;
\det \hat M_0 = \lim_{\e \to 0^+} \e^{-2} \det M_\e ~ .
\quad
\eeq

If this deformation is added to the action, it explicitly
breaks the exact lattice supersymmetry and gauge invariance.
This infrared regulator could be removed in the continuum
limit, say, by taking $\epsilon a \ll N^{-1}$.  Noting that
$L = Na$ is the physical size of the lattice, the
equivalent requirement is that $\epsilon \ll L^{-1}$
be maintained as $a \to 0$, for fixed $L$.  Thus in the
thermodynamic limit ($L \to \infty$), the deformation is
removed.  The parameter $\epsilon$ is a soft
infrared regulating mass, and is quite analogous to the
soft mass $\mu$ introduced by CKKU~\cite{Cohen:2003qw}
in their Eq.~(1.2) to control the bosonic zeromode of
the theory.  
In the same way that the deformation introduced with $\mu$
does not modify the final results of the renormalization
analysis of Section 3.4 of \cite{Cohen:2003qw}, our $\epsilon$
does not modify the result of the quantum continuum limit.
The essence of the argument is that we have introduced a
vertex that will be proportional to the dimensionless
quantity $g_2^2 \epsilon a^3 \ll g_2^2 a^3 / L$, where
$g_2$ is the 2d coupling constant.  Such
contributions to the operator coefficients $C, \hat C$
in Eq.~(3.29) of \cite{Cohen:2003qw} vanish in the thermodynamic
limit. Because the target theory is super-renormalizable,
we are assured that the perturbative power counting
arguments are reliable and the correct continuum limit is
obtained.

In \cite{Giedt:2003vy}, we studied the convergence of
$\det \hat M_\e \to \det \hat M_0$.  We found that the convergence
was rapid and that a reliable estimate for $\det \hat M_0$
can be obtained in this way.  As a check, we computed
the eigenvalues of the undeformed matrix $M$,
using the math package Maple, for
10 random boson configurations.
We found that the product of nonzero eigenvalues
agreed with $\det \hat M_0$ in magnitude and phase
to within at least 5 significant digits.

In the phase-quenched ensemble, we take the probability measure
defined by $|\det M_\e| e^{-S_B}$, with $\e = 10^{-6}$.
Our numerical studies indicate that this $\e$ is more than 
small enough to factor out the ever-present zeros
while leaving the nonzero eigenvalues virtually unchanged.
We have dropped the factor of $\e^{-2}$ that distiguishes
between $\hat M_\e$ and $M_e$.  This is permissible because it cancels
in correlation functions.  In the simulation, this is reflected
in the fact that under a shift in the boson configuration,
$\delta \ln | \det \hat M_\e| = \delta \ln |\det M_\e|$,
since $-2 \ln \e$ is a constant.  The value $\e = 10^{-6}$
renders $M_\e$ rather ill-conditioned, so we must use
a very accurate inversion algorithm (see below).  As an aside,
we have found that approximation of $M_\e^{-1}$ by noisy estimators fails
because the bicongujate gradient algorithm cannot handle
the ill-conditioned matrix.  This is the main obstacle that
we have faced in our attempts to extend our analysis beyond
$N=6$.  Note that for $N=6$, the matrix $M_\e$ is already
$576 \times 576$ with complex entries.

\subsection{Determinant on moduli space}
\label{dms}
We have stated above that the effects of fermions must be
taken into account in order to draw conclusions about
observables of the supersymmetric system.
One way to see this is to consider the fermion determinant in
the case where we set the boson configuration to one of the
points of the undeformed theory moduli space discussed
above in Sec.~\ref{jirr}.  We now show that additional
fermion zeromodes appear in these boson configurations.
This seems reasonable if we regard the lattice action
as the Hamiltonian of a classical statistical system.
Zero action boson configurations correspond to bosons of
zero energy.  By the exact supersymmetry of the undeformed
system, we expect that these zero energy bosonic states
have fermionic partners, also of zero energy.  These fermion
zeromodes will cause the fermion matrix to be more
singular.\footnote{These observations were pointed out
to us by a referee of an earlier version of this 
report \cite{Giedt:2003gf}.}  Indeed, this heuristic
argument seems borne out by the examples below.  However,
we were unable to constuct an explicit proof; presumably
it would involve a study of eigenvalues of the transfer
matrix corresponding to the lattice action.

\subsubsection{$x_\mbf=y_\mbf=0$ non-zeromode branch}  
Here we can choose
\beq
( \xi_{1, \nbf}^\rho , \xi_{2, \nbf}^\rho , \xi_{3, \nbf}^\rho ,
\lambda_\nbf^\rho )^T = ( 0 , 0 , 0 , \lambda_\nbf \delta^{\rho 0} )^T
\eeq
It is easily checked that $M$ acting on this vector
vanishes for any choice of the $N^2$ Grassmann variables
$\lambda_\nbf$.  This gives $N^2$ fermion zeromodes associated
with left multiplication.  We also have $N^2$ fermion
zeromodes associated with right multiplication:
\beq
\( \psi_{1,\mbf}^\mu ~,~ \psi_{2,\mbf}^\mu ~,~
\psi_{3,\mbf}^\mu ~,~ \chi_\mbf^\mu \)
= \( 0 ~,~ 0 ~,~
\psi_\mbf \delta^{0 \mu} ~,~ 0 \)
\eeq
Note that the number of zeromodes, $2N^2$, is the same
as the number of (complex) boson zeromodes found in
Sec.~\ref{oops} above.

Two of these modes are the ever-present zeromodes
discussed in Sec.~\ref{uies} above.  The other $2(N^2-1)$ are zeros
special to this branch of moduli space.
Strictly speaking, the set of configurations that
satisfy $x_\mbf=y_\mbf=0$ is a set of measure
zero in the lattice partition function.  But taking
a small neighborhood of radius $\delta$ about this
subspace---that is, small fluctuations about the
classical configuration---we see that $\det M_\e
\propto \e^2 \delta^{2(N^2-1)}$ in this region, where
the powers of $\delta$ come from the $2(N^2-1)$
fermion approximately-zero-modes that are present
near the $x_\mbf=y_\mbf=0$ branch.  Thus the
inclusion of fermions dramatically changes the
weight of configuration space in the neighborhood of
$x_\mbf=y_\mbf=0$ branch in the integration
measure; it has become totally negligible.
Thus {\it we do not expect the entropic effects discussed
in Sec.~\ref{intp} to persist in the supersymmetric
system.}  As a consequence, we expect to get a
very different value for $\vev{\phi^{x,0}}$.

\subsubsection{$z_\mbf=0$ non-zeromode branch}
First consider the case where $\omega^{k'_1+k'_2} = 1$.
Then set
\beq
( \xi_{1, \nbf}^\rho , \xi_{2, \nbf}^\rho , \xi_{3, \nbf}^\rho ,
\lambda_\nbf^\rho )^T = ( 0 , 0 , \delta^{\rho 0} \xi ,
0 )^T \quad \forall \quad \nbf
\eeq
It is easy to check that this is a fermion zeromode on
this branch, for any choice of $\xi$.  There is also a 
corresponding fermion zeromode of right multiplication:
\beq
\( \psi_{1,\mbf}^\mu ~,~ \psi_{2,\mbf}^\mu ~,~
\psi_{3,\mbf}^\mu ~,~ \chi_\mbf^\mu \)
= \( 0 ~,~ 0 ~,~ 0 ~,~ \delta^{\mu 0} \chi \)
\quad \forall \quad \mbf
\eeq

Now consider the case where $\omega^{k'_1+k'_2} = -1$.
Then set
\beq
( \xi_{1, \nbf}^\rho , \xi_{2, \nbf}^\rho , \xi_{3, \nbf}^\rho ,
\lambda_\nbf^\rho )^T = ( 0 , 0 , \delta^{\rho 0} (-)^{n_1+n_2} \xi ,
0 )^T \quad \forall \quad \nbf
\eeq
It is only slightly more work to check that this is a fermion zeromode on
this branch.  There is also a corresponding fermion zeromode of right
multiplication:
\beq
\( \psi_{1,\mbf}^\mu ~,~ \psi_{2,\mbf}^\mu ~,~
\psi_{3,\mbf}^\mu ~,~ \chi_\mbf^\mu \)
= \( 0 ~,~ 0 ~,~ 0 ~,~ \delta^{\mu 0} (-)^{m_1+m_2} \chi \)
\quad \forall \quad \mbf
\eeq
We note that the suppression due to the fermion determinant is
much smaller, relative to the $x_\mbf=y_\mbf=0$ branch,
in the neighborhood of this branch:
$\det M_\e \propto \e^2 \delta^2$.

\subsection{Determinant distribution in the fully-quenched ensemble}
\label{erew}
Formally, the expectation value of an operator $\Ocal$ in
the phase-quenched ensemble is related to a
ratio of expectation values in the fully-quenched 
({\it f.q.}) ensemble:
\beq
\vev{\Ocal}_{p.q.} = \vev{|\det M| \Ocal}_{f.q.} /
\vev{|\det M|}_{f.q.}
\label{jslr}
\eeq
This is a type of reweighting identity.
However, for this to work in practice, it is necessary that there be a
significant overlap between the phase-quenched ensemble and
the fully-quenched ensemble.  We now show that this is certainly
not the case in the present system, which is just to say
that the fully-quenched expectation values have no bearing
on the system with fermions.  Our main point here is just
to show that the inclusion of the effects of fermions {\it in the
updating process of the Monte Carlo simulation} is
crucial to the study of the supersymmetric system,
a not very surprising result.  We now quantify this
statement by an analysis of the impracticality of \myref{jslr}.

We accumulate $N_s$ points of data in the fully-quenched ensemble and
then bin them according to $\ln |\det M|$.  From the
number of samples $n_i$ in each bin $i$, we obtain an estimate
$w_i \approx n_i/N_s$ of the weight of each bin.  We denote the average
value of $\phi_\mbf^{x,0}$ in each bin by $\vev{\phi_\mbf^{x,0}}_i$.
We denote
\beq
|\det M|_i = \exp \( \hbox{central value of} \; \ln |\det M| \;
\hbox{in bin} \; i \)
\eeq
In principal, as the number of bins is taken to infinity
and the resolution of each bin is taken to be infinitely fine,
the approximations
\beq
\vev{|\det M| \phi_\mbf^{x,0} }_{f.q.} 
& \approx & 
\sum_i w_i |\det M|_i \vev{\phi_\mbf^{x,0}}_i \nnn
\vev{|\det M|}_{f.q.} & \approx &
\sum_i w_i |\det M|_i
\eeq
become exact.
In practice, however, we always have truncation errors from
using a finite number of bins, and systematic errors from
finite resolution.  Furthermore, the statistics in all
of the bins that contribute significantly to these sums
must be large, so that error estimates for the quantites
$w_i$ and $\vev{\phi_\mbf^{x,0}}_i$ are reasonably small.

\begin{figure}
\begin{center}
\includegraphics[height=4.5in,width=3in,angle=90]{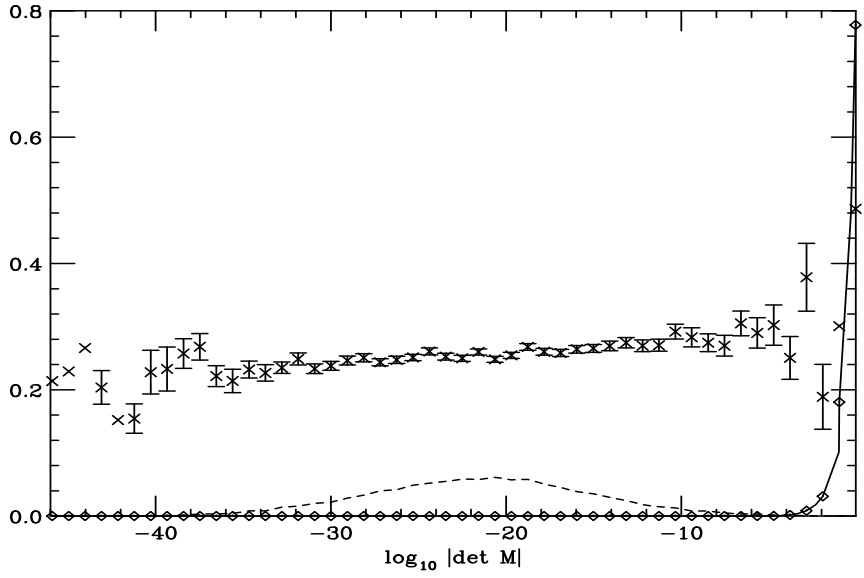}
\end{center}
\caption{The dashed line is $w_i$.  The solid line connects data
for $w_i |\det M|_i$, marked by diamonds.  Data marked by an
``x'' represent $\vev{\phi_\mbf^{x,0}}_i$.  Where error bars are not shown
for the $\vev{\phi_\mbf^{x,0}}_i$ data, 
only 1 or 2 points of data were available, so no estimate of
error could be made.  This plot is for 10000 samples on
a $6 \times 6$ lattice with $\hginv=2$.
\label{nola}}
\end{figure}

In Fig.~\ref{nola} we show $w_i, \vev{\phi_\mbf^{x,0}}_i$
and $w_i |\det M|_i$ as a function of $\log_{10}|\det M|_i$.
It can be seen that the overall weight $w_i |\det M|_i$ is
sharply peaked in a region where $w_i$ is very small. (On the
plot, $w_i$ is indistinguishable from zero in this
region; numerically, it is $\ord{10^{-4}}$.)  
The estimate of $\vev{\phi_\mbf^{x,0}}_i$
in these upper bins is very inaccurate, due to a small number
of samples.  In fact, even for 10000 samples, we were
not able to see the right-hand side of the peak!  Thus
it is impossible to estimate the truncation error that
results from having data for only a finite number of bins.
For the present system, it is not possible to
get a reliable estimate of either of the averages on the right-hand
side of \myref{jslr}.  This is consistent with the results
of Sec.~\ref{dms}:  the fully-quenched ensemble gives its
largest weight to configurations that have practically no
weight in the phase-quenched ensemble.  Thus the overlap
is essentially nil, rendering \myref{jslr} quite useless.
Another conclusion that we can draw from this is: 
expectation values obtained in the fully-quenched ensemble
are not a good indicator, to any degree of approximation,
of expectation values that would be obtained in the
phase-quenched ensemble.

\subsection{Results for the phase-quenched ensemble}
\label{oope}
\begin{figure}
\begin{center}
\includegraphics[height=4.5in,width=3in,angle=90]{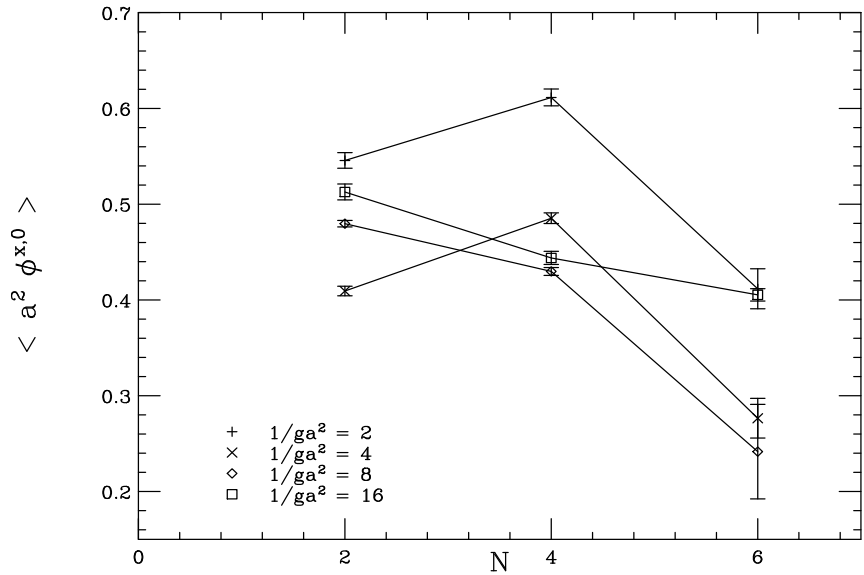}
\end{center}
\caption{Phase-quenched averages (lines to guide the eye). 
\label{lupq}}
\end{figure}

In Fig.~\ref{lupq} $\vev{\hat \phi_\mbf^{x,0}}$ is obtained in the phase-quenched
ensemble; that is, the fermion determinant was taken into
account in Metropolis updates, using the linear variation
\beq
\delta \ln |\det M_\e| = \real \tr [ M_\e^{-1} \delta M_\e ]
\eeq
As usual, we worked at $\e = 10^{-6}$, so that the ever-present
fermion zeromodes factor out.  This renders $M_\e$ ill-conditioned.
To invert the matrix reliably, we have used triangular (LU) decomposition
by Crout's method, which is very accurate.  We have checked that
the inversion is good to 1 part in $10^6$ in our simulations.

\begin{figure}
\begin{center}
\includegraphics[height=4.5in,width=3in,angle=90]{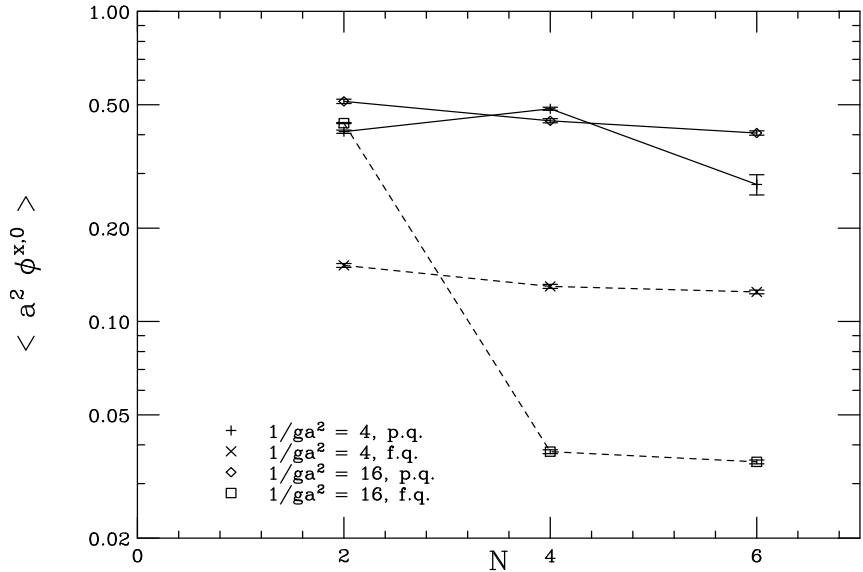}
\end{center}
\caption{A comparison of some of the
averages obtained in phase-quenched (solid lines to
guide the eye) versus
fully-quenched (dashed lines to guide the eye) simulations.
\label{lupqfq}}
\end{figure}

We remind the reader that larger values of $\hginv = 1/g a^2$
correspond to finer lattices, and thus extrapolate to the
continuum limit.  However, to hold $L=Na$ fixed,
successive trajectories must be compared
as $N$ versus $2N$, since $\hginv$ is doubled.  It can be
seen from Fig.~\ref{lupq} that for $N \leq 6$ it is not possible to extrapolate
to the large $L$ behavior.  The best we can say
is the following.
In Fig.~\ref{lupqfq} we provide a comparison between the
phase-quenched and fully-quenched results.  It can
be seen that the effect of the fermion determinant
is dramatic.  For the finest lattice that we have
available, $\hginv = 16$, the phase-quenched results are
more than an order of magnitude larger when we look
at $N=6$.  Furthermore, the phase-quenched results are
always of the same order of magnitude as the classical
estimate \myref{jjus}.

\subsection{Phase reweighting}
\label{pooe}
\begin{figure}
\begin{center}
\includegraphics[height=4.5in,width=3.0in,angle=90]{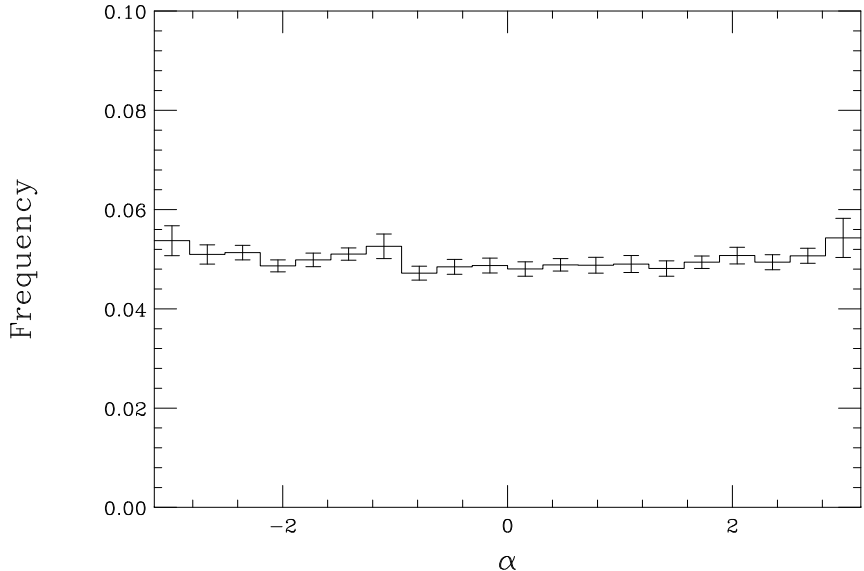}
\end{center}
\caption{Average frequency distribution 
for $\alpha = \arg \det \hat M_0$ in the phase-quenched distribution
for the $2 \times 2$ lattice.
Error bars are determined by variance in bin counts for several
blocks of data. \label{phdb}}
\end{figure}

As has been reported earlier \cite{Giedt:2003vy}, and
mentioned above, the distribution
of $\alpha = \arg \det \hat M_0$ is essentially flat with respect
to the phase-quenched ensemble.  In Fig.~\ref{phdb} we present
results for a $2 \times 2$ lattice.  It can be seen that it
will be quite difficult to obtain a
reliable estimate of $\vev{\exp(i \alpha)}_{p.q.}$.

E.g., we find that for $N=2$, $\hginv=2$, with 1000 samples,
the ``noise-to-signal'' ratio is
\beq
\s_{\vev{\cos \alpha }_{p.q.}} / \vev{\cos \alpha }_{p.q.} =
\s_{\vev{\sin \alpha }_{p.q.}} / \vev{\sin \alpha }_{p.q.} = \ord{1}
\eeq
For $N=4$, $\hginv=2$, with 1000 samples, we find
\beq
\s_{\vev{\cos \alpha }_{p.q.}} / \vev{\cos \alpha }_{p.q.} = 
\s_{\vev{\sin \alpha }_{p.q.}} / \vev{\sin \alpha }_{p.q.} = \ord{10}
\eeq
On general grounds, these noise-to-signal ratios are expected to
get exponentially worse as $N$ is increased.
Similarly discouraging results are found for the other quantities
that need to be estimated in order to make use of the reweighting
identity \myref{tslr}.  Since reweighting does not even work for
small $N$, it looks like a futile approach by which to study
the full theory.

\mys{Conclusions}
\label{s:con}
In the fully-quenched system we have found that 
the classical estimate for the expectation value
$\vev{\phi^{x,0}}$ is not reliable.  We have argued that this is due to 
entropic effects that dominate when the relative
strength of the deformation is scaled to zero.  Thus
we conclude that the fully-quenched system does not
yield the nonsupersymmetric gauge theory in the 
quantum continuum limit.  However, we regard it as
an interesting question whether or not a well-defined
continuum theory is obtained if the relative strength
of the deformation is not scaled to zero.  We have
not investigated this question as it is beyond the
intended scope of this article.

Next we have shown that the fermion determinant tends to vanish
in the neighborhood of the configurations that were argued to
yield the dominating entropic effects.  It follows that the
fully-quenched ensemble will not provide a reliable
guide to the physics of the full, supersymmetric theory.
We have studied $\vev{\phi^{x,0}}$ in the phase-quenched ensemble.
We have found a dramatic increase in the expectation value
that is obtained.  We regard this as encouraging for
the CKKU proposal.  

However, to truly obtain data on the full lattice theory
proposed by CKKU, it is necessary to take into account
the complex phase of the fermion determinant.
We have discussed the notion of
reweighting by the phase of the determinant.  However,
we know from previous work that
the distribution of this phase
in the phase-quenched ensemble is so nearly flat that
it is not practical to obtain reweighted quantities.
As a practical matter, it seems impossible to study the 
continuum limit of the full theory by Monte Carlo methods,
unless there is significant progress in the technique
for the treatment of the complex phase of the fermion
determinant.

In this regard, we remark that the complex phase is not present
in the target continuum theory; so, it may be possible to
address the phase by blocking out the short distance modes
of the lattice theory.  However, this too is beyond the
scope of this report.  Alternatively, it may be possible
to improve on the CKKU construction and eliminate the
complex phase altogether.  We are particularly interested
in the constructions that do not involve a dynamical
lattice spacing such as the one that appears here.
Many attempts to formulate supersymmetric field theories 
on the lattice may be found in the literature (see for example 
\cite{Curci:1986sm,Montvay:1994ze,Fleming:2000fa,Montvay:2001aj,Feo:2002yi} 
and references therein).  In our opinion, the
recent successes in non-gauge models are very encouraging 
\cite{Catterall:2000rv,Catterall:2001wx,Catterall:2001fr}.  
Of particular importance is the understanding
of these models as ``topological'' or {\it Q-exact,} where $Q$ is an
exact supercharge of the lattice system 
\cite{Catterall:2003wd,Catterall:2003xx,Catterall:2003uf}.
Indeed, the CKKU undeformed action can be written in a Q-exact form.
An exciting development has been Sugino's exploitation of this Q-exact 
idea to constuct lattice super-Yang-Mills with {\it compact} gauge fields and
an ordinary (non-dynamical) lattice spacing \cite{Sugino:2003yb}.  
While the Sugino construction has its benefits, he notes
(based on a remark by Y. Shamir)
that these systems also suffer from a vacuum degeneracy
problem that renders the classical continuum limit
ambiguous.  Consequently, Sugino has introduced a non-supersymmetric
deformation in these models that lifts the unwanted
vacua; he demands that the relative strength of
this deformation be scaled to zero in the thermodynamic
limit.  Thus in some respects the Sugino construction
has a feature that is similar to the models of CKKU.
He argues that entropic effects do not destroy the
vacuum selection imposed by his deformation.  We are
currently investigating various aspects
of Sugino's proposal \cite{enm}.

\vspace{15pt}

\noindent
{\bf \Large Acknowledgements}

\vspace{5pt}

\noindent
The author would like to thank Erich Poppitz for comments,
especially in regard to the moduli space of the undeformed
theory.  Thanks are also due to a referee, of an earlier
version of this report \cite{Giedt:2003gf}, who provided 
useful criticism and suggestions.  We thank Mithat \"Unsal
for useful communications.
This work was supported by the National Science and Engineering 
Research Council of Canada and the Ontario 
Premier's Research Excellence Award.

\myappendix

\mys{$\Ncal=4$ moduli space}
Here we establish the well-known solution to \myref{qers}.
One way to see this is as follows \cite{Fayet:1978ig}.  First we note that
$S_0$ reduced to the zero modes, which we write as $S_z$,
takes the form
\beq
S_z &=& \frac{N^2}{g^2} \tr \( \half ([\xdag,x] + [\ydag,y] + [\zdag,z])^2 \right.
\nnn && \left. +2 [x,y][\ydag,\xdag] + 2[y,z][\zdag,\ydag] +2 [z,x][\xdag,\zdag] \)
\label{jhas}
\eeq
Now note that the $U(1)$ parts of $x,y,z$ do not appear and can
take any value.  Thus we can restrict our attention the the $SU(k)$
parts, which we choose to express in terms of Hermitian matrices
$a_p,b_p, \; p=1,2,3$:
\beq
x^c T^c &=& (a_1^c + i b_1^c) T^c = a_1 + i b_1 \nnn
y^c T^c &=& (a_2^c + i b_2^c) T^c = a_2 + i b_2 \nnn
z^c T^c &=& (a_3^c + i b_3^c) T^c = a_3 + i b_3
\label{ksfe}
\eeq
Substitution into \myref{jhas} and a bit of algebra yields
\beq
S_z &=& -\frac{N^2}{g^2} \tr \[ 2 \(\sum_p [a_p,b_p]\)^2
+ \sum_{p,q} \( [a_p, b_q] + [b_p, a_q] \)^2 \right. \nnn
&& \left. + \sum_{p,q} \( [a_p, a_q] - [b_p, b_q] \)^2 \] \nnn
&=& -\frac{N^2}{g^2} \sum_{p,q} \tr \( [a_p, a_q]^2
+ [b_p, b_q]^2 + 2 [a_p, b_q]^2 \)
\eeq
Using positivity arguments quite similar to those above,
one finds that $S_z \geq 0$  and that $S_z = 0$ iff
\beq
[a_p, a_q] = [b_p, b_q] = [a_p, b_q] = 0 , \quad \forall p,q
\label{oiwe}
\eeq
which is nothing other than Eq.~(53) of \cite{Fayet:1978ig}.
Since the matrices are all hermitian and they all commute, 
it is obviously possible to choose a basis which simultaneously
diagonalizes them.  This basis will be related
to the one used in \myref{ksfe} according to 
$T^c \to T'^c = \alpha T^c \alpha^\dagger$,
which is nothing other than the global 
gauge transformations \myref{oore}.

\end{document}